\documentclass[%
 prd,twocolumn,
 nofootinbib,
groupedaddress,
aps,
]{revtex4-1}
\usepackage{amsmath, amssymb}
\usepackage{graphicx,xcolor}
\usepackage{dcolumn}
\usepackage{bm}
\usepackage{hyperref}

\newcommand{\sirfh}{{\tt SIRFH}}

\begin{document}
\title{COVID-19: Nowcasting Reproduction Factors Using Biased Case Testing Data.}

\author{Carlo~R.~Contaldi}
\email{c.contaldi@imperial.ac.uk}
\affiliation
{Headquarters, Standing Joint Command (UK)
}
\affiliation{%
 Blackett Laboratory, Imperial College London, SW7 2AZ, UK}

\date{\today}

\begin{abstract}
Timely estimation of the current value for COVID-19 reproduction factor $R$ has become a key aim of efforts to inform management strategies. $R$ is an important metric used by policy-makers in setting mitigation levels and is also important for accurate modelling of epidemic progression. This brief paper introduces a method for estimating $R$ from biased case testing data. Using testing data, rather than hospitalisation or death data, provides a much earlier metric along the symptomatic progression scale. This can be hugely important when fighting the exponential nature of an epidemic. We develop a practical estimator and apply it to Scottish case testing data to infer a current (20 May 2020) $R$ value of $0.74$ with 95\% confidence interval $[0.48 - 0.86]$.
\end{abstract}


\maketitle

\section{Introduction}
The worldwide COVID-19 pandemic poses a particular challenge for data--driven epidemiological modelling due to the inability of testing regimes to provide a fair, unbiased sample of the infection rate in the general population. Testing has, understandably, been prioritised for essential, key--workers associated with a country's health systems, emergency services, or critical infrastructure. 

Focusing only these sections of the population means testing is a highly biased tracer of the underlying infection rate. The result is that data analysis has focused on death or hospitalisation data as more robust indicators of disease progression \footnote{\url{www.imperial.ac.uk/mrc-global-infectious-disease-analysis/covid-19}} \footnote{\url{www.mrc-bsu.cam.ac.uk/tackling-covid-19/}}
\footnote{\url{www.ox.ac.uk/coronavirus-research}}. The disadvantage is that these provide a time-lagged metric of the disease with respect to the point of infection. The lag between infection and hospitalisation could be as long as two weeks whilst the lag to recovery or death could be an additional two weeks or more (see e.g. \cite{imperial8}). 

As countries move out of initial lockdowns and into recovery phases the focus will be on indicators of any second wave of infections. Testing data could provide a key indicator for rising levels of transmission but the challenge is to account for the biased selection of the population that is being tested due to limited capacity. In this {\sl paper} we introduce a method for estimating current values of $R$ using biased testing data. This comes under a class of methods know as {\sl nowcasting} and are a more useful input to policy--making than forecasts based on model fits using data that lags the point of infection by many days. Since an epidemic is essentially an exponentially unstable system, providing a feedback to mitigate instabilities at the earliest possible time can change the impact by orders of magnitude.

This {\sl paper} is organised as follows. In Section~\ref{sec:sirfh} we introduce a simple extension of SIR models that includes additional compartments. These are used to show how the effect of multiple pathways from infection to death or recovery can be included into the estimate of reproduction factors. Section~\ref{sec:biased} introduces a model for linear biasing of testing data with respect to underlying infections and shows how this can be combined with a linearised approximation of the early stages of the epidemic in order to obtain an estimator for the reproduction factor $R$ as a function of time. In Section~\ref{sec:estimator} we show how the estimator can be applied to practical situations which require proxies for correction factors. We apply the estimator to Scottish, historic case testing data in Section~\ref{sec:scotland}. The application yields an estimate for $R(t)$ in Scotland. We discuss the results in Section~\ref{sec:disc}.

\section{\sirfh\  compartmental model}\label{sec:sirfh}

In order to determine a method for estimating $R$ using biased testing data we use an extension of the Susceptible--Infected--Removed (SIR) \cite{1927RSPSA.115..700K,2000SIAMR..42..599H} compartmental epidemic model. The extension introduces additional compartments that track hospitalisations and hospital--based fatalities. The \sirfh\  model has been used to fit epidemic parameters to UK hospital and death data in aide to the military response to the current outbreak \cite{jmc_model}.  A more complete model would include latent compartment and non--hospital based fatalities but as we shall see, for the purpose of this estimate, these are not strictly necessary. 

The model consists of a set of coupled ordinary differential equations for the following compartmental break down of a population; Susceptible ($S$), Infected ($I$), Recovered ($R$), Fatalities ($F$), and Hospitalised ($H$). The total population is a conserved number $N = S+I+R+F+H$. The variables are linked via a set of coupling rates which may themselves have additional time-dependence through an external deterministic mechanism,
\begin{align}\label{eq:sirfh}
\frac{dS}{dt} &= - \beta \frac{S}{N}I\,, \ \ \frac{dI}{dt} = \beta \frac{S}{N}I - (\gamma +\alpha)I\,,\nonumber\\
\frac{dR}{dt} &= \gamma I + \zeta H\,, \ \ \frac{dH}{dt} =  \alpha I - (\zeta+\mu)H\,, \\
\frac{dF}{dt} &= \mu H\,,\nonumber
\end{align}
where $t$ is the independent variable (in units of days). Integrating the equations leads to a time evolution for all compartments. Initial conditions are required for integration and usually consist of a seed number of initial infections $I_0$. These also determine the initial number of susceptible individuals $S_0=N-I_0$. All other variables have vanishing initial conditions.

The rate variables are to be interpreted as {\sl effective} rates which may be susceptible to a combination of external factors not directly modelled by the set of differential equations. The model does not include natural mortality or birth processes that would change the underlying total for the population. This simplification is equivalent to assuming that the duration of the epidemic is short compared to the timescale on which the total population changes in the absence of excess deaths.

In Table~\ref{tab:rates} we summarise the physical interpretation of the five coupling rates. The rate $\beta$ is often related to a basic reproduction factor $R_0$ for the disease through $\beta = R_0\gamma$. The reproduction factor describes the average number of new infections caused by an infected individual during the timescale $1/\gamma$ for which the individual is assumed to be infectious. All rates included in the model may themselves be time-dependent with their evolution determine by mechanisms external to the model. $\beta$, in particular, is susceptible to mitigation strategies such as social distancing and lockdowns. In this work we assume that only $\beta$ is time-dependent via $R_0=R(t)$.

\begin{table}[ht]%
    \centering
    \caption{Interpretation of the model rates controlling the movement of the population between model compartments in the SIRFH model (\ref{eq:sirfh})}\label{tab:rates}
    \begin{tabular}{l|cl}
   Parameter && Interpretation\\ 
    \hline\hline
    $\beta$&& Susceptible $\to$ Infected\\
        $\alpha$&& Infected $\to$ Hospital\\
        $\gamma$&&  Infected $\to$ Recovered\\
        $\mu$&& Hospital $\to$ Fatality\\
        $\zeta$&& Hospital $\to$ Recovered\\
    \hline
    \end{tabular}
\end{table}%

\section{Biased, linearised system}\label{sec:biased}

During the initial stages of the epidemic, when $S\approx N(1+\epsilon)$ with $\epsilon \ll 1$, the system (\ref{eq:sirfh}) can be linearised and the equation for $dI/dt$ decoupled to obtain
\begin{equation}\label{eq:rate}
    \frac{d\log I}{dt} = \gamma R (1+\epsilon) - (\gamma +\alpha)\,.
\end{equation}
$\epsilon$ is a perturbation variable which only becomes large when the system departs from the linear regime. In terms of the decoupled system, it can be understood as a correction term accounting for the non--linear coupling between $I$ and $S$ compartments. We can invert (\ref{eq:rate}) to obtain an equation for an estimate $\tilde R(t)$ as a function of the rate at which the logarithm of the number of infections is changing
\begin{equation}
    \tilde R(t) \approx \frac{1}{1+\epsilon}\left[1  + \frac{1}{\gamma}\left( \frac{d \ln I}{dt} + \alpha\right)\right]\,.
\end{equation}
Therefore, if we could establish $\gamma$, $\alpha$, and the rate $d\ln I/dt$, we could determine the instantaneous reproduction factor $R$ if the decoupled, linearised approximation holds. This estimate would be accurate in the early stages of the epidemic when still far from the herd immunity fixed point.

In the absence of a direct measurement for $d\ln I/dt$ we consider how biased testing can be used to evaluate $R(t)$. We introduce a linear bias factor $b(t)$ \cite{1984ApJ...284L...9K} and define
\begin{equation}\label{eq:bias}
P(t) = \frac{1}{N} N_T(t) \, b(t) \, I(t)\,,
\end{equation}
where $P$ is the number of test--confirmed cases at time $t$ that resulted from $N_T$ tests. Both the bias factor and $N_T(t)$ modulate the positive counts with respect to the underlying infection rate $I(t)/N$ but the effect of any changing test capacity can be removed by considering only the {\sl positive fraction} of test results $p(t)\equiv P(t)/N_T(t)$ such that
\begin{equation}
    p(t)= \frac{1}{N}b(t)\,I(t)\,.
\end{equation}
Considering the derivative of $p$ we have
\begin{equation}
    \frac{d \ln I}{dt} = \frac{d \ln p}{dt} - \frac{d \ln b}{dt}\,.
\end{equation}
This allows us to relate $R$ to the rates of change in the logarithm of $p$ and linear bias factor $b$
\begin{equation}\label{eq:R}
    \tilde R(t) = \frac{1}{1+\epsilon}+ \frac{1}{\gamma(1+\epsilon)}\left(\frac{d \ln p}{dt} - \frac{d \ln b}{dt}+\alpha\right)\,.
\end{equation}
This result shows that, since $R$ is related to the rate at which the infection is progressing, it is not sensitive to the biasing itself but to the derivative of the logarithm of bias. If the selection of the population being tested remains relatively constant the term $d\ln b/dt$ will be small and a measurement of $p(t)$ can be used to estimate $R(t)$. The rate $\alpha$ incorporates the rate at which individuals are being removed from the infected compartment through a channel other than recovery. In the \sirfh\ model this is the rate at which infected are hospitalised \footnote{We are assuming there is no nosocomial infection.} and thus removed from the infection chain. It could also describe the rate at which individuals are removed through isolation. Unless the hospitalisation or isolation fraction is comparable to $I(t)/N$ this term will be small. $\alpha < 0.005$ for the current outbreak \cite{jmc_model}. 

\begin{figure}[t]
    \centering
    \includegraphics[width=\linewidth]{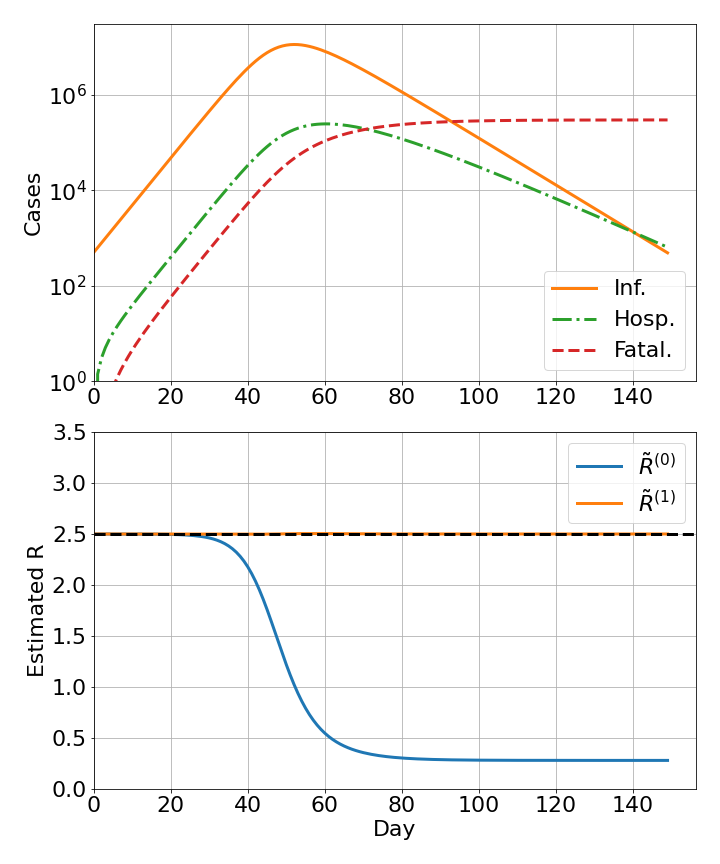}
    \caption{Epidemic curves (top) for a constant $R=2.5$ model with $1/\gamma = 6$ days, $\zeta = 0.05$, $\alpha = 0.0026$, $\mu=0.034$, and 5000 seed infections. The estimate $\tilde R$ (bottom) is shown for both zeroth and first order in $epsilon$. The total population $N$ is 50 million in this model and $b$ is assumed to be constant.}
    \label{fig:model}
\end{figure}

Fig.~\ref{fig:model} shows the application of (\ref{eq:R}) to a model with constant $R=2.5$. The linear bias is assumed to be constant in this example such that $d\ln p/dt = d\ln I/dt$. We use $\epsilon = S/N-1$ and since this is obtained from a full integration of the non-linear system we recover $R$ exactly at first order in $\epsilon$. 

This example is an unmitigated scenario where the disease progresses rapidly through the full infection and reaches herd immunity for the particular value of $R$. The zeroth order estimate $\tilde R^{(0)}$ diverges from the true value as soon as $\epsilon\sim1$ close to the peak of infections. A zeroth order estimate would therefore be of little value in such a scenario. Below we show how some simple refinement can improve on this.

\begin{figure}[t]
    \centering
    \includegraphics[width=\linewidth]{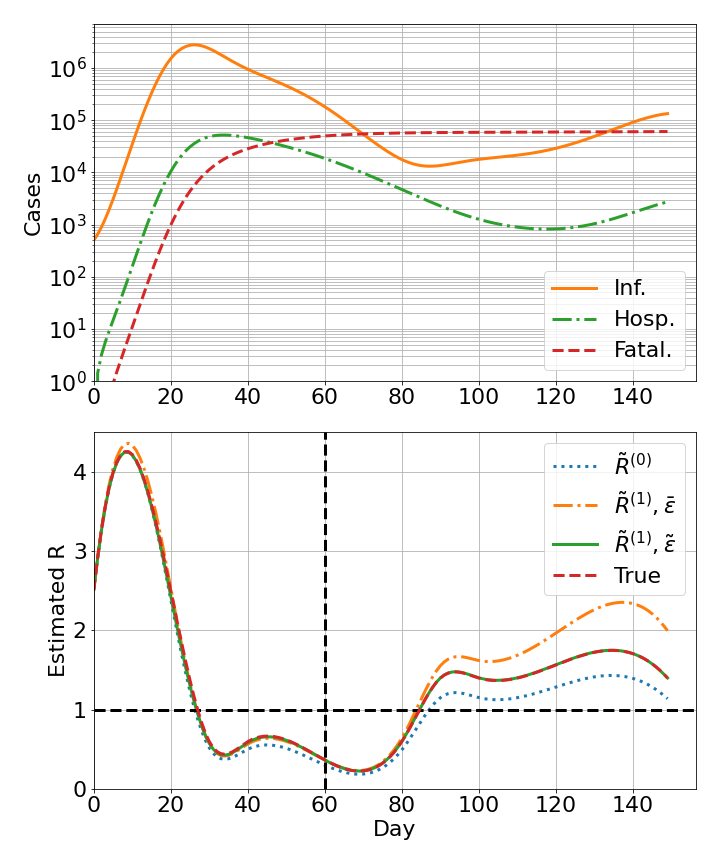}
    \caption{Epidemic curves (top) for a time-dependent $R(t)$ scenario where mitigation is in place. All other rates are same as in the previous scenario. The estimated reproduction factor (bottom) is shown for successive refinements of the correction factor. We assume perfect knowledge of $\theta(t_f)$  at $t_f$ = day 60 (vertical dashed line) but its time-dependence is unknown. The correction $\tilde \epsilon$ using $c(t)$ as a proxy for $\epsilon(t)$ recovers the true $R(t)$ accurately. Despite being normalised around $t_f=60$ this correction works well even beyond the point where the true $R$ increases beyond 1.}
    \label{fig:model2}
\end{figure}

\section{Practical estimation of the reproduction factor}\label{sec:estimator}

In practical situations $b$ is not constant and we do not know $\epsilon$, $\gamma$, and $\alpha$ exactly. In the following we assume $alpha$ is sufficiently small and can be neglected. This will be true for any epidemic where the hospitalisation rate is small compare to the basic transmission rate. 

A distribution for $\gamma$ will be know with some degree of precision from studies of the serial interval of the disease \cite{imperial13}. We will introduce a distribution in the application below. 

Maintaining a constant selection of the population in measuring $p$ will also mitigate against the term $d\ln b/dt$. Alternatively, if $b$ changes in rapid stages, for example through rapid changes in testing regime, it will lead to sharp peaks in  $d\ln b/dt$ which will have limited effects on longer--term trends in $\tilde R$. Here, we focus on a proxy for the correction $\epsilon$. 

A crude approximation for $\epsilon$ can be obtained if we have access to an estimate of the fraction of the population that has been infected by time $t_f$ \footnote{We neglect compartments $H$ and $F$ here.}
\begin{equation}
    \theta(t_f) = \frac{I(t_f)+R(t_f)}{N}\,.
\end{equation}
A linear approximation for $\epsilon$ could be employed
\begin{equation}\label{eq:beps}
    \bar \epsilon(t) = -\frac{\theta(t_f)}{t_f}t\,.
\end{equation}

Assuming we observe the curve $p(t)$ we can integrate it to obtain its normalised cumulative
\begin{equation}
    c(t) = \frac{1}{C}\int^{t}_0 dt\, p(t)\,,
\end{equation}
with 
\begin{equation}
    C = \int^{t_f}_0 dt\, p(t)\,.
\end{equation}
A better proxy for $\epsilon$ would then be
\begin{equation}\label{eq:teps}
    \tilde \epsilon (t) = - \theta(t_f) \,c(t)\,.
\end{equation}

Any approximation for the correction term $\epsilon$ will fail once the number of infected becomes comparable to $N$ but in epidemics where transmission is suppressed using external mitigation and $S$ remains close to $N$ the proxies can be used effectively.

We show an example in Fig.~\ref{fig:model2}. $R$ is now time-dependent with a suppression (lockdown) phase with $R<1$ and a later phase where $R>1$. We use a constant bias factor of 20 to approximate $p(t)$, although this normalisation does not affect the estimate of $R.$

We assume an estimate of $\theta$ is obtained on day 60, during the lockdown. On that day $\theta = 17\%$. This value is used to normalise the approximations $\bar \epsilon(t)$ (\ref{eq:beps}) and $\tilde \epsilon(t)$ (\ref{eq:teps}). The estimate using the correction $\tilde \epsilon$ recovers the true $R(t)$ very accurately even beyond the point $t_f$ and where $R>1$.

\section{Application to Scottish Testing Data}\label{sec:scotland}

Scotland is the only UK nation whose government publishes testing data that includes positive and negative testing numbers\footnote{\url{https://www.gov.scot/publications/coronavirus-covid-19-trends-in-daily-data/}, accessed 23 May.}. The historical record is for the whole of Scotland.

\begin{figure}[t]
    \centering
    \includegraphics[trim=0 100 5 10,clip,width=\linewidth]{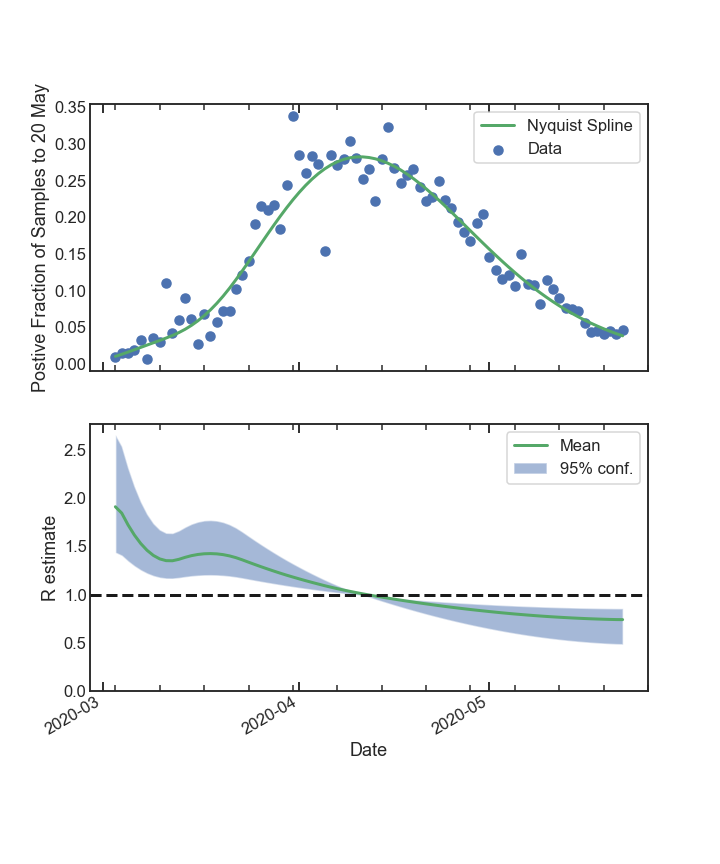}
    \caption{Scottish testing data (top) with a Nyquist limited spline used as an estimate of $p(t)$ and its cumulative function $c(t)$. The $R$ estimate (bottom) shows the result given by the mean value for the serial interval $1/\gamma$ and the 95\% ranges. The estimate assumes 5\% of the Scottish population has been infected so far.}
    \label{fig:scot}
\end{figure}

The results of the estimate is shown in Fig.\ref{fig:scot}. We take the daily positive fraction of tests as the data and fit a univariate spline to $\ln p$ with a number of knots limited to $N_k=L/14$ where $L$ is the length of the set in days (the ''Nyquist'' spline). This ensures that any systematic, weekly modulation due to reporting is smoothed out whilst retaining as much of the long--term trend as possible. We use the smoothed estimate as $p(t)$ and calculate its cumulative function $c(t)$. We assume $\theta$=5\% of the Scottish population has been infected so far \footnote{To refine this one could marginalise over a range in $\theta$.}. 

We assume the infectiousness timescale (serial interval) $1/\gamma$ is a random variable following a Gamma distribution of shape parameter 6.5 and scale parameter 0.62 \cite{imperial13}. We sample the distribution to find the 2.5\%, 50\%, and  97.5\% percentiles and use these values to evaluate a median $\tilde R$ and 95\% confidence intervals around it.

This preliminary analysis indicates that the reproduction factor for the whole of Scotland is $R=0.74$ with 95\% confidence interval $[0.48 - 0.86]$. A key assumption underlying this result is that the Scottish outbreak is no where near heard immunity level which would be closer to the range of 60\% to 70\% as the infected fraction of the total population.

\section{Discussion}\label{sec:disc}

The method discussed here could be used to analyse any regional outbreak where the total number of infected individuals is small compared to the total population. The key requirement for this analysis is the ability to determine the positive fraction of samples tested. This data is not openly available for other UK nations or for Scotland at sub-national level. 

A key advantage of this method is that it uses data from case testing. This is the largest data set upstream of case hospitalisations that is a direct tracer, albeit biased, of the infection. As such it offers a valuable insight for planning purposes and for policy--making. Considering the exponential increase in cases when $R>1$ every day gained in obtaining indications of rising infection rates will result in a large, positive impact for any mitigation strategy. 

Given the volume of data it should be feasible to determine $R$ at sub-national resolution. This could help in setting up systems for ''smart'' lockdowns where mitigation strategies are implemented locally in order to minimise economic and health impact of intervention in areas where transmission rates are not high.

The large volume of data can also be used in mitigating effects of any testing regime changes. For example, samples used could be selected to maintain the bias as constant as possible. 

\acknowledgements 
We acknowledge many useful discussions with Professors Arttu Rajantie and Toby Wiseman. The author would also like to acknowledge the many individuals in the Information Manoeuvre Group (IMG) HQ SJC(UK) whose extraordinary effort to understand the epidemic inspired and enabled this work. The author also acknowledges the large and rapidly evolving body of literature on COVID-19 which they have not been able to adequately survey. As such they are grateful for any indication of related work by readers.

\bibliographystyle{apsrev}
\bibliography{refs.bib}


\end{document}